\documentclass{article}

\begin{document}

\title{A comment on gr-qc/9901053~\footnote{Work supported
             by CONICOR, CONICET and Se.CyT, UNC}}

\author{{\sc Mirta S. Iriondo}\thanks{Partially supported by AIT,
C\'ordoba,
    Argentina. }, {\sc Enzo O. Leguizam\'on} \thanks{Fellow of
    Se.CyT-UNC.} \\
 and   {\sc Oscar A. Reula}
  \thanks{Member of CONICET.}\\
  {\small FaMAF, Medina Allende y Haya de la Torre,}\\
  {\small Ciudad Universitaria, 5000 C\'ordoba, Argentina}\\
}
 
\maketitle

In eprint gr-qc/9901053 Gen Yoneda and Hisa-aki Shinkai have
made several claims disputing our results in \cite{Iriondo1}, and
\cite{Iriondo2}. We show here that these claims are not correct.

The authors begun mentioning three points by which they think our
discussion of the topic is not clear, so we begin clarifying them, for this
shall be useful in making our point on the correctness of our approach:

1.- {\bf On definitions of symmetric-hyperbolic systems:} In the modern
    notation of pseudo-differential equations the definition of
    hyperbolicity is made in terms of the principal part of the symbol
    of the operator, where the symbol is obtained by replacing the
    occurrence of derivatives by $- i k_a$, with $k_a$ a real
    co-vector, thus obtaining a matrix in the cotangent bundle of the
    base space. That is, 
    $P(u(x^j),\frac{\partial}{\partial x^i}) \to p(u(x^j),k_i)$ via
    $\frac{\partial}{\partial x^i} \to - i k_i$.
    Incidentally, we do not comment, as stated by the authors
    at the end of the first section of Appendix C, the case of $k_a$
    complex, we only comment the case of the {\sl norm} of 
    $k_a$, $k := q^{ab}k_ak_b$, being complex, which is the
    case when the metric is complex, we always consider $k_a$ real.
    
    Symmetric hyperbolicity requires the principal part of the symbol
    of a first order quasilinear system
    to be such that there exists a symmetric, positive definite
    bilinear form $h = h(u)$ such that 
    $h(u)p(u,k) + p(u,k)^{\dagger} h(u)= 0$,
    that is, anti-hermiticity of the symbol with respect to the
    scalar product defined by $h$.

    So our definition is the standard one and is {\it equivalent} to the one
    used by the authors,
    so the results are the same, and the modification of the system to
    make it symmetric-hyperbolic adding terms proportional to the
    constraints are the same we performed in our first paper.

    Contrary to what the authors assert in the last paragraph of page
    2, symmetric hyperbolicity of a system implies the Cauchy problem
    for that systems is well posed, {\sl even for the quasilinear case}.
    The proof of this fact is
    not new and appears in most modern textbooks in the theory of PDE's.

2.- {\bf On reality conditions:} 
    It is clear from equation (2.38a) in \cite{ash} that if 
    $D_a$ \hbox{${}_{{}_{{}_{\widetilde{}}}}$\kern-.5em\it N} $=0$
    then our modified system also preserves the triad reality
    condition, and there is no inconsistency. But what we claim is that 
    our system does not need of the triad
    reality condition to be symmetric hyperbolic. 
    This is important for the choice of foliation implied by the
    constancy of the densitized lapse is too restrictive for successful 
    future applications in numerical simulations. 
    The fact that the triad reality condition is not needed is
    explained in detail in our second paper, for there it is displayed
    in all details the scalar
    product (or what is equivalent $h(u)$) in which the symbol is
    anti-hermitian (and therefore symmetric-hyperbolic) with respect
    to. 
    The matrix $h(u)$ is hermitian and
    positive definite by construction, this is a standard procedure in
    the theory of PDE's from the point of view of pseudo-differential
    operators. 
    We {\bf do not do} what the authors claim we do in our work,
    which they call a
    ``proposal'', namely to rotate the triad to anti-hermiticity and then use
    the equations for the rotated system. This is not needed.    
    What we do is to find a ``rotated'' scalar product in which the
    system is symmetry-hyperbolic, that is to find a more general symmetrizer. 
    What {\bf we do need} is the metric reality conditions, for
    otherwise the ``rotation'' to anti-hermiticity of the triad can not be
    done. They are treated in detail in the second
    paper.

3.- {\bf The characteristic structure of the resulting system:} This
    structure is not needed for asserting symmetric-hyperbolicity, so
    we did not include it in our first paper, for it was a letter, we
    did include it in our second paper, where we explicitly
    diagonalized the system, and where we also displayed the
    characteristic structure of the constraint evolution equations,
    since they might be useful for evolution, and since the
    eigenvectors structure is remarkable simple.

From the second point above, using standard definitions, and
displaying explicitly the scalar product used for
symmetric-hyperbolicity it is clear that our work is correct,
furthermore the calculations in the work we are commenting are the
same than those we have previously done.

\end{document}